\documentclass[10pt]{article}

\usepackage{amstext,amsmath,amssymb,amsfonts,bbm,slashed}
\usepackage[latin1]{inputenc}
\usepackage{epsfig}
\usepackage{hyperref}
\usepackage{amsthm}
\usepackage{subfigure}
\usepackage{color}
\usepackage{multirow}

\newtheorem{remark}{Remark}
\newtheorem{definition}{Definition}

\newtheorem{proposition}{Proposition}

\newcommand{\bea}{\begin{eqnarray}}
\newcommand{\eea}{\end{eqnarray}}
\newcommand{\beq}{\begin{equation}}
\newcommand{\eeq}{\end{equation}}

\newcommand{\bpro}{\begin{pro}}
	\newcommand{\epro}{\end{pro}}
\newcommand{\blem}{\begin{lem}}
	\newcommand{\elem}{\end{lem}}
\newcommand{\bdfn}{\begin{dfn}}
	\newcommand{\edfn}{\end{dfn}}
\newcommand{\bcor}{\begin{cor}}
	\newcommand{\ecor}{\end{cor}}
\newcommand{\bthm}{\begin{thm}}
	\newcommand{\ethm}{\end{thm}}
\newcommand{\bex}{\begin{ex}}
	\newcommand{\eex}{\end{ex}}
\newcommand{\brmk}{\begin{rmk}}
	\newcommand{\ermk}{\end{rmk}}
\newcommand{\bpr}{\begin{pr}}
	\newcommand{\epr}{\end{pr}}

\makeatletter

\begin{document}

	\begin{center}
		
		{\LARGE\bf   Hamiltonian Dynamics
		  for the Kepler  Problem in a  Deformed Phase Space }

		\vspace{15pt}
		
		{\large   Mahouton Norbert Hounkonnou and  Mahougnon Justin Landalidji 
		}
		
		\vspace{15pt}
		
		{ 
		International Chair of Mathematical Physics
		               and Applications  (ICMPA-UNESCO Chair)\\
		              University of Abomey-Calavi, 072 B.P. 50 Cotonou, Republic of Benin\\
		         E-mails: { norbert.hounkonnou@cipma.uac.bj with copy to
		                hounkonnou@yahoo.fr} \\
		         E-mails: { landalidjijustin@yahoo.fr}   
		   }

	\end{center}
	\vspace{10pt}
	\begin{abstract}
		This work addresses the Hamiltonian dynamics of the Kepler problem
		in a deformed phase space, by considering the equatorial orbit. The
		recursion operators are constructed and used to compute the
		integrals of motion.  The same investigation is performed with the
		introduction of the Laplace-Runge-Lenz vector.  The existence of
		quasi-bi-Hamiltonian structures is also elucidated. Related
		properties are studied.
		
		
	\end{abstract}
	
	\section{Introduction}\label{sec_1}
	In 1601, Kepler obtained a detailed set of observations of the motion
	of the planet Mars from the Danish astronomer Tycho Brahe \cite{Brac}.
	From his analysis of these data, Kepler determined that the path of
	Mars is an ellipse, with the sun located at a focal point, and that
	the radius vector from the sun to the planet sweeps out equal areas in
	equal times.  The direct problem was to determinate the nature of the
	force required to maintain elliptical motion about a focal force
	center.  This direct problem remained unsolved until after 1679, when
	Newton determined the functional dependence on distance of the force
	required to sustain such an elliptical path of Mars about the sun as a
	center of force located at a focal point of the ellipse.
	
	Building on Newton's description of the nature and universality of the
	gravitational force, scientists of the eighteenth century shifted
	their interest almost exclusively from direct to inverse problems.
	They used the combined gravitational forces of the sun and the other
	planets to predict and explain perturbations in the conic paths of
	planets and comets.  That interest continued through the nineteenth
	and twentieth centuries, and today scientists still concentrate upon
	the inverse problem rather than the direct one.
	
	In particular, in the last few decades there was a renewed interest in
	the Kepler problem as one of completely integrable Hamiltonian systems
	(IHS), the concept of which goes back to Liouville in 1897 \cite{lio}
	and Poincar\'{e} in 1899 \cite{po}.  Loosely speaking, IHS are
	dynamical systems admitting a Hamiltonian description, and possessing
	sufficiently many constants of motion.  Many of these systems are
	Hamiltonian systems with respect to two compatible symplectic
	structures \cite{mag1,gel,vil1,fil1} leading
	to a geometrical interpretation of the so-called recursion operator
	\cite{lax}.  The theory of integrable Hamiltonian systems, based on
	the use of the Nijenhuis torsion, is a part of the geometry of a
	particular class of manifolds, called Poisson-Nijenhuis manifolds
	\cite{mag2}.  In 1992, Marmo and Vilasi \cite{mar} constructed a
	recursion operator for the Kepler dynamics, and obtained related
	constants of motion.
	
	From the Magri works \cite{mag1,mag2}, it is known that the
	eigenvalues of the recursion operator of bi-Hamiltonian systems form a
	set of pairwise Poisson-commuting invariants \cite{bro4}. It is,
	however, worth noticing that two kinds of difficulties often arise,
	while investigating these systems: $(i)$ Firstly, it is in general
	very difficult to give locally an explicit second Hamiltonian
	structure for a given integrable Hamiltonian system \cite{rav} even if
	it is theoretically always possible in the neighborhood of a regular
	point of the Hamiltonian \cite{bro2}; $(ii)$ Secondly, the global or
	semi-local existence of such structures implies very strong conditions
	which are rarely satisfied \cite{bro,fer}.
	
	In 1996, R. Brouzet {\it et al.} defined a weaker notion under the
	name of {\it quasi-bi-Hamiltonian system} (QBHS) which relaxes these
	two difficulties for two degrees of freedom.  In 2000, G. Sparano {\it
		et al} constructed recursion operator for the Kepler dynamics, in
	the non-commutative case using the so-called Delauney action-angle
	coordinates \cite{spa}.  Further, in 2013, Hosokawa and Takeuchi
	\cite{kh} solved the same problem, but using the Runge-Lenz-Pauli
	vector, and got new constants of motion.  A bi-Hamiltonian formulation
	for a Kepler problem was also studied with Delaunay-type variables
	\cite{gri}.  In 2016, J. F. Cari\~{n}ena {\it et al.} \cite{car}
	investigated some properties of the Kepler problem related to the
	existence of quasi-bi-Hamiltonian structures.  In this work, we
	investigate the Kepler dynamics in a deformed phase space.
	
	The paper is organized as follows.  In Section \ref{sec_3}, we present
	the considered deformed phase space.  In Section \ref{sec_4}, we
	define, in action-angle coordinates, the deformed Hamiltonian
	function, symplectic form and vector field describing the Kepler
	dynamics.  In Section \ref{sec_6}, we construct recursion operators,
	and compute the associated integrals of motion. In Section
	\ref{sec_7}, we give an alternative Hamiltonian description for the
	dynamical systems and obtain associated recursion operators in a
	non resonant case. In Section \ref{sec_8}, we study the existence of
	quasi-bi-Hamiltonian structure for the considered Kepler dynamics. In
	Section \ref{sec_9}, we end with some concluding remarks.
	
	\section{Deformed phase space and Kepler Hamiltonian}\label{sec_3}
	
	Let $\mathbb{R}_{0}^{3} = \mathbb{R}^{3} \backslash \{0,0,0\}$ be the
	configuration manifold $\mathcal{Q},$ and
	$\mathcal{T}^{\ast}\mathcal{Q} = \mathcal{Q} \times \mathbb{R}^{3}$ be
	the cotangent bundle with the local coordinates $(q,p).$ The cotangent
	bundle $\mathcal{T}^{\ast}\mathcal{Q}$ has a natural symplectic
	structure $\omega$ which, in local coordinates, is given by
	
	\[ \omega = \sum_{i = 1}^{3} dq^{i} \wedge dp_{i}.\] Since $\omega$ is
	non-degenerate, it induces the map $\Lambda$:
	$\mathcal{T}^{\ast}\mathcal{Q} \longrightarrow \mathcal{T}\mathcal{Q}$
	defined by
	
	\[ \Lambda = \sum_{i = 1}^{3} \dfrac{\partial}{\partial q^{i}}\wedge
	\dfrac{\partial}{\partial p_{i}},\] where $\mathcal{T}\mathcal{Q}$
	is the tangent bundle. The map $\Lambda$ is called the bivector field
	\cite{vil2} and used to construct the Hamiltonian vector field $X_{f}$
	of a Hamiltonian function $f$ by the relation
	\begin{equation}\label{Kvec}
	X_{f} = \Lambda df.
	\end{equation}  
	The phase space deformation is here understood by replacing the usual
	product with the $\gamma-$star product, (also known as the Moyal
	product law) between two arbitrary functions of position and momentum
	\cite{vak,ma,khos1} :
	\begin{equation} \label{Eq_3_1} (f\ast_{\gamma}g) (q,p) =
	f(q_{i},p_{i})  \exp{\bigg(\dfrac{1}{2}\gamma^{ab}\overleftarrow{\partial}_{a}\overrightarrow{\partial_{b}}\bigg)}
	g(q_{j},p_{j})\Bigg|_{(q_{i},p_{i})= (q_{j},p_{j})},
	\end{equation}
	where
	\begin{equation} \label{Eq_3_2} \gamma^{ab} =
	\begin{pmatrix}
	\varTheta^{ij} & \delta^{i}_{j}  \\
	- \delta^{i}_{j}  & 0 \\
	\end{pmatrix},
	\end{equation}
	$\varTheta$ is an antisymmetric $n \times n$ matrix inducing the
	deformation in the coordinates.  Without loss of generality, we
	restrict our study to the first two terms of the $\ast_{\gamma}$
	deformed Poisson bracket expansion to obtain:
	
	\begin{equation}
	\{ f,g\}_{\gamma} = \varTheta^{ij}\dfrac{\partial{f}}{\partial{q^{i}}}\dfrac{\partial{g}}{\partial{q^{j}}} + \bigg(\dfrac{\partial{f}}{\partial{q^{i}}}\dfrac{\partial{g}}{\partial{p_{i}}} - \dfrac{\partial{f}}{\partial{p_{i}}}\dfrac{\partial{g}}{\partial{q^{i}}}  \bigg),
	\end{equation}
	giving
	\begin{equation}\label{Eq_3_4}
	\{q^{i},q^{j}\}_{\gamma} = \varTheta^{ij}, \ \{q^{i},p_{j}\}_{\gamma} = \delta^{i}_{j}, \  \{p_{i},p_{j}\}_{\gamma} =0.
	\end{equation}
	The Kepler Hamiltonian in $\mathcal{T}^{\ast}\mathcal{Q}$ takes the
	form:
	\begin{equation}
	H = \dfrac{p_{i}p^{i}}{2m} + V(r),
	\end{equation}
	yielding the Hamilton's equations:
	\begin{equation} \label{eq2} \dot{q}^{i} := \{ q^{i} ,H\}_{\gamma}=
	\dfrac{p^{i}}{m} +
	\varTheta^{ij}\dfrac{\partial{V(r)}}{\partial{q^{j}}} ; \quad
	\dot{p}_{i} := \{ p_{i},H\}_{\gamma}= -
	\dfrac{\partial{V(r)}}{\partial{q^{i}}}
	\end{equation}
	and the following correction to the Newton second law \cite{ro1} :
	\begin{equation}
	m\ddot{q}^{i} = -\dfrac{q^{i}}{r}\dfrac{k}{r^{2}} + m\varepsilon^{ijk}\dot{q}^{j}\varOmega^{k} + m \varepsilon^{ijk}q^{j}\dot{\varOmega}^{k},
	\end{equation}
	where the deformation parameter
	$\varTheta^{ij} = \varepsilon^{ijk}\alpha^{k},$ and the angular
	velocity
	\[ \varOmega^{i} = \dfrac{k}{r^{3}}\alpha^{i}, i= 1,2,3. \] Setting
	the deformation parameter $\alpha^{i} = \delta^{i3}\alpha$ transforms
	$H$ into
	
	\begin{equation}\label{eq4'}
	H = \dfrac{m}{2}\bigg[ (\dot{q}^{1} - q^{2}\varOmega)^{2} + (\dot{q}^{2} + q^{1}\varOmega)^{2} + (\dot{q}^{3})^{2}\bigg] - \dfrac{k}{r},
	\end{equation}
	which is reduced to:
	\begin{equation}\label{Ham}
	H = \dfrac{p_{r}^{2}}{2m} + \dfrac{p_{\varphi_{\alpha}}^{2}}{2mr^{2}} - \dfrac{k}{r}
	\end{equation}
	in spherical coordinates $(r,\upsilon,\varphi)$, and equatorial orbit
	corresponding to $\upsilon = \dfrac{\pi}{2},$ where
	$ p_{r} = m\dot{r}$ and
	$ p_{\varphi_{\alpha}} = mr^{2} \dot{\varphi}_{\alpha},$ with
	$\dot{\varphi}_{\alpha} = (\dot{\varphi} + \varOmega)$ and
	$\varphi_{\alpha} = (\varphi + \varOmega t) \in (0, 2\pi ).$
	
	Equation \eqref{eq4'} encodes the information on the phase space deformation through  $ \varOmega,$ which depends on the deformation parameter $\alpha.$ However, it can  evidently be interpreted as equivalent
	to
	the Hamiltonian for a charged particle in a homogeneous, independent of
	time,  magnetic field along $z$ axis, and the central Newtonian gravitational
	field in the usual commutative space.
	
	Now considering the coordinate system
	$(r,\varphi_{\alpha},p_{r},p_{\varphi_{\alpha}}),$ and using
	\eqref{Kvec}, we get the following Hamiltonian vector field:
	\begin{equation}\label{Hamvk}
	X_{H} = \dfrac{1}{m}\Bigg[p_{r}\dfrac{\partial}{\partial r} - \dfrac{1}{r^{3}}\bigg( - p_{\varphi_{\alpha}}^{2}  + mkr \bigg)\dfrac{\partial}{\partial p_{r}}  + \dfrac{p_{\varphi_{\alpha}}}{m r^{2}}\dfrac{\partial}{\partial\varphi_{\alpha}}\Bigg].  
	\end{equation}
	
	\section{Hamiltonian system in the action-angle
		coordinates} \label{sec_4}
	
	The Hamiltonian function \eqref{Ham} does not explicitly depend on the
	time.  Then, setting $V= W - Et, $ it is possible to find a complete
	integral for the equation of motion by using the method of variable
	separation:
	\begin{equation}
	W = W_{r}(r) + W_{\varphi_{\alpha}} (\varphi_{\alpha}).
	\end{equation}
	In this case, the Hamilton-Jacobi equation \cite{arn} is reduced to
	\begin{equation}
	E= \dfrac{1}{2m}\bigg( \dfrac{\partial{W}}{\partial{r}}\bigg)^{2} + \dfrac{1}{2mr^{2}}\bigg( \dfrac{\partial{W}}{\partial{\varphi_{\alpha}}}\bigg)^{2} - \dfrac{k}{r},
	\end{equation}
	leading to the following set of equations:
	\begin{equation*}
	\begin{cases}
	\bigg( \dfrac{dW_{\varphi_{\alpha}} (\varphi_{\alpha})}{d\varphi_{\alpha}}\bigg)^{2}= D_{\varphi_{\alpha}}^{2} \\
	-r^{2}\bigg(\dfrac{dW_{r}(r)}{dr}\bigg)^{2} + 2mr^{2}E + 2mrk =
	D_{\varphi_{\alpha}}^{2},
	\end{cases}
	\end{equation*}
	where $D_{\varphi_{\alpha}}$ is constant.  In the compact case
	\cite{vil2}, characterized by $E < 0,$ we can introduce the action
	variables \cite{abra} $J_{r}$ and $J_{\varphi_{\alpha}}$ such that:
	\begin{equation*}
	\begin{cases}
	J_{\varphi_{\alpha}}= \dfrac{1}{2\pi}\oint \dfrac{dW_{\varphi_{\alpha}}  (\varphi_{\alpha})}{d\varphi_{\alpha}}d\varphi_{\alpha} \\
	J_{r} =\dfrac{1}{2\pi} \oint \dfrac{dW_{r}(r)}{dr}dr.
	\end{cases}
	\end{equation*}
	Using the method of residue \cite{abl,vil2}, we get
	\begin{equation*}
	J_{r} = - p_{\varphi_{\alpha}} + \dfrac{mk}{\sqrt{-2mE}}, \ \ D_{\varphi_{\alpha}} = p_{\varphi_{\alpha}},
	\end{equation*}
	and the integrable system \cite{Bog}: {\small
		\begin{equation}\label{lio3}
		\begin{cases}
		\dot{J}_{i}= 0
		\\
		\dot{\varphi}^{i}= \dfrac{\partial H}{\partial J_{i}},
		\end{cases}
		\Rightarrow 
		\begin{cases}{ll}
		J_{1}= J_{r}; \ J_{2}= J_{\varphi_{\alpha}}
		\\
		\varphi^{1}= \dfrac{mk^{2}}{(J_{1} + J_{2} )^{3}}t;\ \varphi^{2}
		= \dfrac{mk^{2}}{(J_{1} + J_{2} )^{3}}t, \ \varphi^{1}(0) =
		\varphi^{2} (0) = 0.
		\end{cases}
		\end{equation}}

	\begin{proposition} 
		In action-angle coordinates $(J,\varphi),$ the Hamiltonian $H,$ the
		symplectic form $ \omega$ , and the Hamiltonian vector field $X_{H}$
		are, respectively:
		\begin{align} \label{Bil0} & H = E = - \dfrac{mk^{2}}{2(J_{1} +
			J_{2})^{2}}; \ \omega = \sum_{h = 1}^{2}dJ_{h}\wedge
		d\varphi^{h}, \\ \label{Bil2} \mbox{and} \cr X_{H} &= \{H,.\} :=
		\dfrac{mk^{2}}{(J_{1} + J_{2} )^{3}}\bigg(
		\dfrac{\partial}{\partial{\varphi^{1}}} +
		\dfrac{\partial}{\partial{\varphi^{2}}}\bigg),
		\end{align}
		where $\{. ,.\}$ is the usual Poisson bracket.
	\end{proposition}
	
	\section{Recursion operators } \label{sec_6} Let us define a $2$-form
	$\omega_{1}$ and a vector field $\Delta,$
	\begin{equation}
	\omega_{1}: = \sum_{h,k = 1}^{2} S_{h}^{k}dJ_{k}\wedge d\varphi^{h} = \sum_{h = 1}^{2}d\lambda_{h}\wedge d\varphi^{h} , \quad \Delta := \lambda_{h} \dfrac{\partial}{\partial{J_{h}}},
	\end{equation} 
	where $S =
	\begin{pmatrix}
	J_{1} & J_{2}    \\
	J_{2} & J_{1}
	\end{pmatrix},\quad
	\begin{cases}
	\lambda_{1} = \dfrac{1}{2}\bigg( J_{1}^{2} + J_{2}^{2} \bigg) \\
	\lambda_{2} = J_{2}J_{1},
	\end{cases}$ such that $\omega_{1}$ is the Lie derivative of the
	symplectic form $\omega$ in \eqref{Bil0} with respect to the vector
	field $\Delta,$ $i.e.$:
	\[\mathcal{L}_{\varDelta}\omega = \omega_{1}. \]
	
	The vector field $\Delta$ generates a sequence of finitely many
	(Abelian) symmetries according to the following scheme:
	\[ X_{i + 1} := [X_{i}, \Delta]_{\mu}= \dfrac{2}{\mu}( X_{i}(\Delta) -
	\Delta(X_{i})),\] where $\mu = 3 - i, \ i=0,1,2 $ and
	$X_{0} = X_{H}$ in \eqref{Bil2}.  The $X_i$'s are given by
	\begin{align} \label{X_1} & X_{0} = \dfrac{mk^{2}}{(J_{1} + J_{2}
		)^{3}}\bigg( \dfrac{\partial}{\partial{\varphi^{1}}} +
	\dfrac{\partial}{\partial{\varphi^{2}}}\bigg), \quad X_{1} =
	\dfrac{mk^{2}}{(J_{1} + J_{2})^{2}}\bigg(
	\dfrac{\partial}{\partial{\varphi^{1}}} +
	\dfrac{\partial}{\partial{\varphi^{2}}}\bigg), \\ \label{X_2} &
	X_{2} = \dfrac{mk^{2}}{(J_{1} + J_{2} )}\bigg(
	\dfrac{\partial}{\partial{\varphi^{1}}} +
	\dfrac{\partial}{\partial{\varphi^{2}}}\bigg), \quad X_{3} =
	mk^{2}\bigg( \dfrac{\partial}{\partial{\varphi^{1}}} +
	\dfrac{\partial}{\partial{\varphi^{2}}}\bigg),
	\end{align}
	are:
	\begin{itemize}
		\item[(i)] in involution, \textit{i.e.},
		\begin{equation} [X_{h}, X_{k}]_{\mu} = 0, \quad h,k = 0,1,2,3, \ \
		\mu = 1,2,3.
		\end{equation}
		\item[(ii)] Hamiltonian vector fields, $i. e.$, can be expressed as:
		\begin{equation}
		X_{i} = \{H_{i}, .\} = \{H_{i + 1}, .\}_{_{1}}, \quad \ i= 0, 1, 2, 
		\end{equation}
	\end{itemize}
	with respect to the Poisson bracket $ \{ .,.\}_{_{1}}$ defined by
	\begin{equation}
	\{ f,g\}_{_{1}} := \sum_{h,k = 1}^{n}(S^{-1})^{h}_{k} \bigg(\dfrac{\partial{f}}{\partial{J_{k}}}\dfrac{\partial{g}}{\partial{\varphi^{h}}} - \dfrac{\partial{f}}{\partial{\varphi^{h}}}\dfrac{\partial{g}}{\partial{J_{k}}}\bigg),
	\end{equation}
	where \[ S^{-1} = \begin{pmatrix}
	\dfrac{J_{1}}{(J_{1} - J_{2}) (J_{1} + J_{2} )}& \dfrac{- J_{2} }{(J_{1} - J_{2}) (J_{1} + J_{2})} \\
	\dfrac{- J_{2} }{(J_{1} - J_{2}) (J_{1} + J_{2} )} & \dfrac{ J_{1}
	}{(J_{1} - J_{2}) (J_{1} + J_{2})}
	\end{pmatrix},\] and
	\begin{equation}
	H_{0} = \dfrac{- mk^{2}}{2(J_{1} + J_{2} )^{2}}, \ H_{1} = \dfrac{- mk^{2}}{(J_{1} + J_{2} )}, \
	H_{2} =  mk^{2} \ln(J_{1} + J_{2} ), \ H_{3} = mk^{2}(J_{1} + J_{2} ). 
	\end{equation}
	\begin{proposition}
		The recursion operator for the Kepler dynamics in the action-angle
		coordinates $(J,\varphi)$ is given by:
		\[
		T = \sum_{h,k}(S)_{k}^{h}\bigg(
		\dfrac{\partial}{\partial{J_{h}}}\otimes dJ_{k} +
		\dfrac{\partial}{\partial{\varphi_{h}}}\otimes d\varphi_{k}\bigg),
		\text{ where } \ S =
		\begin{pmatrix}
		J_{1} & J_{2}    \\
		J_{2} & J_{1}
		\end{pmatrix},\] $\mathcal{L}_{X_{l}}T = 0, \ (l = 0,1,2,3),$ and
		the Nijenhuis torsion vanishes, $i.e.,$ \\
		$(\mathcal{N}_{T})^{h}_{ij}:=T^{k}_{i}\dfrac{\partial{T^{h}_{j}}}{\partial{J^{k}}}
		- T^{k}_{j}\dfrac{\partial{T^{h}_{i}}}{\partial{J^{k}}} +
		T^{h}_{k}\dfrac{\partial{T^{k}_{i}}}{\partial{J^{j}}} -
		T^{h}_{k}\dfrac{\partial{T^{k}_{j}}}{\partial{J^{i}}}= 0, \ (i,j,k,
		h = 1,2).$
	\end{proposition}
	Consider the constants of motion \cite{ro2},
	\[H \ \text{in} \ \eqref{Bil0}, \ M = mr^{2} (\dot{\varphi} + 2
	\varOmega), \ \text{ and } \ L_{\alpha} = M + m\alpha H,\] $i.e.,$
	\begin{equation}
	\{H,M\}= 0, \ \{H,L_{\alpha}\}= 0 ,\ \{M,L_{\alpha}\} = 0.
	\end{equation}
	Then, there exist functions $\phi_{1}$, $\phi_{2}$ satisfying
	\begin{equation} \label{Bil1} \omega' = d\xi_{1}\wedge d\phi_{1} +
	d\xi_{2}\wedge d\phi_{2},
	\end{equation}
	such that the equations of motion in the coordinate system
	$(\xi, \phi)$ are
	\begin{equation}\label{fix1}
	\begin{cases}
	\dot{\xi}_{i}= 0
	\\
	\dot{\phi}^{i}= \dfrac{\partial H'}{\partial \xi_{i}},
	\end{cases}
	\Rightarrow 
	\begin{cases}
	\xi_{i}= cst
	\\
	\phi^{i}= \dfrac{\partial H'}{\partial \xi_{i}}t , \quad
	\phi^{i}(0) = 0,\;\;i \in \{ 1,2 \},
	\end{cases}
	\end{equation}
	where $ \xi_{1} = L_{\alpha},$ $ \xi_{2} = M,$
	$H'= \dfrac{1}{m\alpha} (\xi_{1} - \xi_{2}).$ We get the relationships
	\begin{align}\label{fix}
	& J_{1} = - \xi_{2} +   \varpi + \sqrt{\dfrac{m^{2}\alpha k^{2}}{2(\xi_{2} -\xi_{1})}}; \quad J_{2} = \xi_{2} - \varpi , \\
	\label{fix3}
	&\varphi^{1}=  \dfrac{2\sqrt{2\alpha}}{m\alpha k}(\xi_{2} - \xi_{1} )^{3/2}\phi^{1}; \quad
	\varphi^{2}= -  \dfrac{2\sqrt{2\alpha}}{m\alpha k}(\xi_{2} - \xi_{1} )^{3/2}\phi^{2},   
	\end{align}  
	where $\varpi = mr^{2}\varOmega,$ $\xi_{2}> \xi_{1}>0, \alpha > 0 .$
	Finally, we arrive at:
	\begin{proposition}
		In the coordinate system $(\xi,\phi),$ the Hamiltonian function
		$H',$ the symplectic form $ \omega'$, the Hamiltonian vector field
		$X'_{H}$ and the recursion operator $T'$ are, respectively:
		\begin{align}
		H' &= \dfrac{1}{m\alpha} (\xi_{1} - \xi_{2}) ; \
		\omega'= \sum_{h = 1}^{2}d\xi_{h}\wedge d\phi^{h}; \  
		X'_{H'} = \dfrac{1}{m\alpha} \bigg(\dfrac{\partial}{\partial{\phi_{1}}} - \dfrac{\partial}{\partial{\phi_{2}}}   \bigg) \\
		T' &= \sum_{i = 1}^{2} R_{i}\Bigg(\dfrac{\partial}{\partial{\xi_{i}}} \otimes d\xi_{i} + \dfrac{\partial}{\partial{\phi_{i}}} \otimes d\phi_{i}\Bigg), \text{ where} \ R = 
		\begin{pmatrix}
		\xi_{1} & 0 \\
		0 & \xi_{2}
		\end{pmatrix}.
		\end{align}
	\end{proposition}
	Two interesting cases deserve investigation:
	\begin{itemize}
		\item[I)] Introduce the Laplace-Runge-Lenz (LRL) vector $A$ given by
		\cite{gol}
		\begin{equation}
		A = p \times L - mk\dfrac{q}{r},
		\end{equation}
		where $p$ is the momentum vector, $q$ is the position vector of the
		particle of mass $m$, and $L$ is the angle momentum vector,
		$L = q\times p$ \cite{t}. We obtain:
		\begin{equation} \label{agu} L_{1} = 0; \ L_{2} = 0; \ L_{3} =
		mr^{2} \dot{\varphi}_{\alpha} = p_{\varphi_{\alpha}}.
		\end{equation}
		\begin{equation}
		A_{1} = C \sin\beta  + D \cos\beta; \ 
		A_{2} = C \cos\beta - D\sin\beta; \
		A_{3} = 0,
		\end{equation}
		\begin{align}
		\{A_{1}, H\} &:= \Bigg(\dfrac{\partial{A_{1}}}{\partial{r}}\dfrac{\partial{H}}{\partial{p_{r}}} - \dfrac{\partial{A_{1}}}{\partial{p_{r}}}\dfrac{\partial{H}}{\partial{r}}\Bigg) + \Bigg(\dfrac{\partial{A_{1}}}{\partial{\varphi_{\alpha}}}\dfrac{\partial{H}}{\partial{p_{\varphi_{\alpha}}}} - \dfrac{\partial{A_{1}}}{\partial{p_{\varphi_{\alpha}}}}\dfrac{\partial{H}}{\partial \varphi_{\alpha}}\Bigg) \cr 
		&\hphantom{:=\Bigg(\dfrac{\partial{A_{1}}}{\partial{r}}\dfrac{\partial{H}}{\partial{p_{r}}} - \dfrac{\partial{A_{1}}}{\partial{p_{r}}}\dfrac{\partial{H}}{\partial{r}}\Bigg)} \cr
		&= \dfrac{3 k \alpha p_{r}}{mr^{4}} (D \sin\beta - B\cos\beta), \\   
		\{A_{2}, H\}  & = \dfrac{3 k \alpha p_{r}}{mr^{4}} (B \sin\beta - D\cos\beta),
		\end{align}  
		where
		\[ C = -p_{r}p_{\varphi_{\alpha}}\cos\varphi_{\alpha} +
		\dfrac{p^{2}_{\varphi_{\alpha}}}{r} \sin\varphi_{\alpha} -
		mk\sin\varphi_{\alpha}\] and
		\[ D = p_{r}p_{\varphi_{\alpha}}\sin \varphi_{\alpha} +
		\dfrac{p^{2}_{\varphi_{\alpha}}}{r} \cos\varphi_{\alpha} -
		mk\cos\varphi_{\alpha}, \text{ and } \ \beta =\varOmega t.\]
		\begin{remark}
			We have:
			\begin{itemize}
				\item[(i)] The $A_{i}, \;\; i= 1, 2, 3,$ commute with the
				Hamiltonian $H$ in \eqref{Ham}, $i.e.,$ $\{A_{i},H\} = 0,$ if
				\begin{equation}
				\beta = \dfrac{\pi}{4}; \quad \dfrac{p_{r}p_{\varphi_{\alpha}}}{ \dfrac{p^{2}_{\varphi_{\alpha}}}{r} - mk}= -\cot(\beta + \varphi_{\alpha}),\;\;(\beta + \varphi_{\alpha}) \in (0, \pi). 
				\end{equation}
				\item[(ii)]
				$ \{A_{1}, A_{2}\} = ( - 2 mH + \dfrac{3 k \alpha
					p_{r}}{r^{4}})p_{\varphi_{\alpha}}, \ \{A_{1}, L_{3}\} =
				A_{2},$ and $\{A_{2}, L_{3}\} = A_{1}.$
				\item[(iii)] Setting $L_{3} =A_{3}$ and
				$ p^{2}_{\varphi_{\alpha}} = r[2mk -r + p_{r} (3\varOmega -
				rp_{r})] \equiv A^{2}_{3},$ then, the
				$A_{i} {\textquoteright s}$ generate an $su(2)$ Lie algebra,
				$i.e.,$ $\{A_{i}, A_{j}\} = \varepsilon_{ijl}A_{l}.$
			\end{itemize}
		\end{remark}
		\item[II)] Consider a scaled Runge-Lenz-Pauli vector $\Gamma,$ defined
		on the domain
		$\{(q,p)\in \mathcal{T^{\ast}}(\mathbb{R}^{3}\backslash \{0,0,0\})|
		H(q,p) < 0\}$ by
		\begin{equation} \label{rlp} \Gamma= \dfrac{1}{\sqrt{-2mH}}A,
		\end{equation}
		where $H$ is the Hamiltonian function given in \eqref{Ham}. The
		components $\Gamma_{i}$ are:
		\begin{equation}
		\Gamma_{1} =\dfrac{1}{\sqrt{-2mH}}( C \sin\beta  + D \cos\beta); \ 
		\Gamma_{2} = \dfrac{1}{\sqrt{-2mH}}( C \cos\beta - D\sin\beta); \
		\Gamma_{3} = 0,
		\end{equation}
		with
		\begin{equation}
		|\Gamma|^{2} = - \dfrac{mk^{2}}{2 H} + L^{2}_{3}.
		\end{equation}
		The quantities $H$, $|\Gamma|^{2},$ and $L_{3}$ are in involution,
		$i.e.,$
		\[
		\{|\Gamma|^{2},L_{3}\} = 0 ,\ \ \{|\Gamma|^{2},H\} = 0, \ \
		\{L_{3},H\} = 0.
		\]
	\end{itemize}
	Putting $\pi_{1} = |\Gamma|^{2}$ and
	$\pi_{2} = \ p_{\varphi_{\alpha}},$ the equations of motion in the
	$(\pi, \chi)$ system become:
	\begin{equation}\label{fix4}
	\begin{cases}
	\dot{\pi}_{i}= 0
	\\
	\dot{\chi}^{i}= \dfrac{\partial H''}{\partial \pi_{i}},\;\;H'' =
	\dfrac{m k^{2}}{2(\pi_{2}^{2} -\pi_{1})}
	\end{cases}
	\Rightarrow 
	\begin{cases}
	\pi_{i}= cst, \;\; i= 1, 2.
	\\
	\chi^{i}= \dfrac{\partial H''}{\partial \xi_{i}}t +
	\chi^{i}(0),\quad \chi^{i}(0)= 0.
	\end{cases}
	\end{equation}	  
	The relationships between $(J,\varphi)$ and $(\pi, \chi)$ are deduced
	as:
	\begin{equation}\label{PI1}
	J_{1} = - \pi_{1} +    \sqrt{\pi_{1} - \pi_{2}^{2}}; \ J_{2} = \pi_{2}; \ \chi^{1} =  \dfrac{1}{(J_{1} + J_{2} )}\varphi^{1}; \ \chi^{2} = - \dfrac{J_{2} }{(J_{1} + J_{2} )} \varphi^{2}.
	\end{equation}
	Finally, we get:
	\begin{proposition}
		In the coordinate system $(\pi,\chi),$ the Hamiltonian function
		$H'',$ the symplectic form $ \omega''$, the Hamiltonian vector field
		$X''_{H''}$ and the recursion operator $T''$ are given as follows:
		\begin{align}
		& H'' = \dfrac{m k^{2}}{2(\pi_{2}^{2} -\pi_{1})}; \
		\omega'' = \sum_{h = 1}^{2}d\pi_{h}\wedge d\chi^{h}; \
		X''_{H''} = \dfrac{mk^{2}}{2(\pi_{2}^{2} - \pi_{1})^{2}} \bigg(\dfrac{\partial}{\partial{\chi^{1}}} - 2\pi_{2} \dfrac{\partial}{\partial{\chi^{2}}}\bigg) \\
		& T'' =  \sum_{i = 1}^{2} F_{i}\Bigg(\dfrac{\partial}{\partial{\pi_{i}}} \otimes d\pi_{i} + \dfrac{\partial}{\partial{\chi^{i}}} \otimes d\chi^{i}\Bigg), \text{ where } \  F = 
		\begin{pmatrix}
		\pi_{1} & 0 \\
		0 & \pi_{2}
		\end{pmatrix}.
		\end{align}
	\end{proposition}
	\section{Alternative Hamiltonian description} \label{sec_7}
	
	Let
	\begin{equation} \label{AD} \Upsilon = J_{1}X_{1} + J_{2}X_{2}
	\end{equation}
	be a dynamical system on the manifold $\mathcal{T}^{\ast}\mathcal{Q}$,
	with $ X_{1}$ and $ X_{2}$ obtained in \eqref{X_1} and
	\eqref{X_2}. The relation \eqref{AD} can be rewritten as:
	\begin{equation} \label{AD1} \Upsilon = \nu_{a}X^{a} +\nu_{e}X^{e},
	\end{equation}
	where
	\begin{align} \label{nu}
	&\nu_{a} = -2 H_{a}, \ \nu_{e} = H_{e}; \ H_{a} = J_{1} H_{0}, \  H_{e} = J_{2} H_{1}; \ X^{a} = \dfrac{\partial}{\partial\Phi^{a}},\ X^{e}= \dfrac{\partial}{\partial\Phi^{e}}, \\
	& \dfrac{\partial}{\partial\Phi^{a}} = \bigg(
	\dfrac{\partial}{\partial\varphi^{1}} +
	\dfrac{\partial}{\partial\varphi^{2}} \bigg), \quad
	\dfrac{\partial}{\partial\Phi^{e}} = - \bigg(
	\dfrac{\partial}{\partial\varphi^{1}} +
	\dfrac{\partial}{\partial\varphi^{2}} \bigg).
	\end{align}
	The vector fields $X^{a}, X^{e}$ and the $C^{\infty}-$functions
	$ H_{a}, H_{e}$ satisfy the following properties:
	\begin{equation} [X^{i}, X^{j}] = 0; \quad \mathcal{L}_{X^{i}} H_{i} =
	0, \ \ i,j \in \{ a,e\}.
	\end{equation} 
	Let $\mathcal{ N}$ be an open dense submanifold of
	$\mathcal{T}^{\ast}\mathcal{Q}$ on which $\Upsilon$ is explicitly
	integrable such that:
	\begin{equation} \label{c1} X^{a}\wedge X^{e}\neq 0; \quad dH_{a}
	\wedge dH_{e} \neq 0.
	\end{equation} 
	Now, considering the coordinate system $(H,\Phi)$ with $\Phi^{i},$
	$i \in \{ a,e\},$ which are closed differential 1-forms, the equations
	of motion of $\Upsilon$ are given by
	
	\begin{equation}
	\dot{\Phi}^{a}= -2H_{a}; \
	\dot{\Phi}^{e} = H_{e}; \
	\dot{H}_{a} = 0; \
	\dot{H}_{e} = 0,  
	\end{equation}
	with the functions $H_{a}$ and $H_{e}$ obeying the condition
	\eqref{c1}. We can construct a closed $2$-form, for $i \in \{a,e\}$,
	\begin{equation} \label{om} \tilde{\omega} = \sum_{i}
	df^{i}(H_{i})\wedge d\Phi^{i},
	\end{equation}
	which is non degenerate as long as $df^{a} \wedge df^{e}\neq 0 $, and
	\begin{equation} \label{c2} \iota_{_{X^{i}}}\tilde{\omega} = -df^{i};
	\quad \iota_{_{\Upsilon}}\tilde{\omega} = - \sum_{i} \nu_{i} df^{i};
	\quad \sum_{i} d\nu_{i}\wedge df^{i} = 0.
	\end{equation}  
	Notice that \eqref{c2} is a necessary condition for
	$\iota_{_{\Upsilon}}\tilde{\omega} $ to be exact, $i.e.,$ it is
	closed.  Since $d\nu_{a} \wedge d\nu_{e} \neq 0$, the solutions of
	\eqref{c2} are given by linear functions \cite{gl}
	\begin{equation}
	f^{i} = \sum_{j} L^{ij}\nu_{j}, \quad i, j \in \{a,e\}, \text{ where } \  L=       \begin{pmatrix}
	-1/2 & 0 \\
	0 & 1
	\end{pmatrix}.
	\end{equation}
	Then, we get :
	\begin{equation} \label{fi} f^{a}= - \dfrac{1}{2}\nu_{a}; \quad f^{e}=
	\nu_{e}.
	\end{equation}
	From \eqref{nu} and \eqref{fi}, we can rewrite \eqref{om} in the new
	coordinate system $(\nu, \Phi)$ as:
	
	\begin{equation} \label{om1} \tilde{\omega} = \sum_{i}
	df^{i}(\nu_{i})\wedge d\Phi^{i}, \ \ i \in \{a,e\},
	\end{equation}
	leading to the following form:
	\begin{equation} \label{om2} \tilde{\omega} = -
	\dfrac{1}{2}d\nu_{a}\wedge d\Phi^{a} + d\nu_{e}\wedge d\Phi^{e}.
	\end{equation}
	The corresponding Hamiltonian description for $\Upsilon$ is given with
	the following quadratic Hamiltonian function
	\begin{equation} \label{Ha2} \tilde{ H } = - \dfrac{1}{4}\nu_{a}^{2} +
	\dfrac{1}{2}\nu_{e}^{2}.
	\end{equation}
	In addition, from \cite{vil2} other symplectic structures of the form
	\eqref{om1} can be constructed, in which any $f_{i} $ depending only
	on the corresponding frequency $\nu_{i}, \ i \in \{a,e\}$, will be
	admissible as long as $\tilde{\omega}_{b}, \ b \in \{1,...,n\}, $ is
	non-degenerate, $i.e.$, as long as $df^{a} \wedge df^{e} \neq 0.$ From
	above, putting:
	\begin{equation} \label{fi1} f^{a}= \nu_{a}; \ f^{e}= \nu_{e} \ \text{
		and } f^{a}= \nu_{a}^{2}; \ f^{e}= \nu_{e}^{2}
	\end{equation}
	we obtain, respectively:
	\begin{equation} \label{om3} \tilde{\omega}_{1} = d\nu_{a}\wedge
	d\Phi^{a} + d\nu_{e}\wedge d\Phi^{e} \ \text{ and } \
	\tilde{\omega}_{2} = 2\nu_{a}d\nu_{a}\wedge d\Phi^{a} +
	2\nu_{e}d\nu_{e}\wedge d\Phi^{e}.
	\end{equation} 
	Then, the $(1,1)-$tensor field
	$\mathcal{ T} = \tilde{\omega}_{2} \circ \tilde{\omega}_{1}^{-1}$ is
	constructed, taking the form
	\begin{equation}
	\mathcal{ T} = \mathcal{ T}_{1} + \mathcal{ T}_{2},  
	\end{equation}
	where
	\begin{equation}
	\mathcal{ T}_{1} =2 \nu_{a} \bigg(\dfrac{\partial}{\partial\nu_{a}}\otimes d\nu_{a} + \dfrac{\partial}{\partial \Phi^{a}}\otimes d\Phi^{a} \bigg) \ 
	\text{ and } \ 
	\mathcal{ T}_{2} = 2 \nu_{e} \bigg(\dfrac{\partial}{\partial\nu_{e}}\otimes d\nu_{e} + \dfrac{\partial}{\partial \Phi^{e}}\otimes d\Phi^{e} \bigg).  
	\end{equation}
	Finally, basing on \cite{t,ta,ta2}, $\mathcal{ T}_{1}$
	and $\mathcal{ T}_{2}$ are recursion operators for the dynamical
	system $\Upsilon$. Hence, $\mathcal{ T}$ is also a recursion operator
	for the dynamical system $\Upsilon$ as a sum of two recursion
	operators.
	
	\section{Quasi-bi-Hamiltonian structures} \label{sec_8} Basing on
	\cite{bro} and \cite{car}, in this part, we investigate the recursion
	operators for quasi-bi-Hamiltonian structures.
	\begin{definition}
		A Hamiltonian vector field $Y$ on a symplectic manifold
		$(\mathcal{M}, \omega) $ is called quasi-bi-Hamiltonian if there
		exist another symplectic structure $\omega_{1}$, and a
		nowhere-vanishing function $g$, such that $g Y$ is a Hamiltonian
		vector field with respect to $ \omega_{1},$ $i.e.,$
		\begin{equation}
		\iota_{_{Y}}\omega_{0} = -dH_{0}; \quad
		\iota_{_{gY}}\omega_{1} =\iota_{_{Y}}(g\omega_{1})  = -dH_{1},
		\end{equation}	
		where $H_{0}$ and $H_{1}$ are integrals of motion for the
		Hamiltonian vector field $Y.$ $g\omega_{1}$ is not closed in
		general.
	\end{definition}
	A consequence of this definition is that the pair
	$(\omega_{0}, \omega_{1})$ determines a $(1, 1)$-tensor field $T$
	defined as $T := \hat{\omega}_{0}^{-1}\circ \hat{\omega}_{1},$ that
	is, $\omega_{0}(Y, X) = \omega_{1}(T Y, X ),$ where $X, \ Y$ are two
	Hamiltonian vector fields, and $\hat{\omega} := \iota_{_{Y}}\omega.$
	In the action-angle coordinates $(J,\varphi),$ the decomposition of
	the symplectic form
	
	$ \omega' = \omega'_{1} + \omega'_{2}$,
	\begin{align}
	\omega'_{1} &= dJ_{1}\wedge d\varphi^{1} - \bigg( \dfrac{2 (J_{1} + J_{2})^{3}}{m^{2} k^{2} \alpha} + 1   \bigg) dJ_{2}\wedge d\varphi^{2}  \\
	\omega'_{2} &=    - dJ_{1}\wedge d\varphi^{2} + \bigg( \dfrac{2 (J_{1} + J_{2})^{3}}{m^{2} k^{2} \alpha} + 1   \bigg) dJ_{2}\wedge d\varphi^{1}, 
	\end{align}
	shows that:
	\begin{itemize}
		\item[(i)] $\omega'_{1}$ and $\omega'_{2}$ are not closed, $i.e.$,
		$d\omega'_{1} \neq 0, d\omega'_{2} \neq 0$, where $d$ is the
		exterior derivative. So, $\omega'_{1}$ and $\omega'_{2}$ are not
		symplectic.
		\item[(ii)]
		$ \iota_{_{X_{H}}}\omega'_{1} = -dh'_{1}, \
		\iota_{_{X_{H}}}\omega'_{2} = -dh'_{2}, \ \text{ where } \ h'_{1} =
		-h'_{2}= - \dfrac{2 J_{2}}{m\alpha}.  $
		\item[(iii)] The functions $ h'_{1}$ and $ h'_{2}$ are first integrals
		of $X_{H},$ $i.e., \ X_{H}( h'_{1}) = X_{H}( h'_{2}) = 0.$
	\end{itemize}
	
	\begin{proposition}
		The Hamiltonian vector field $X_{H}$ is quasi-bi-Hamiltonian with
		respect to the two $2$-forms $(\omega, \omega'_{1}).$ Idem for
		$(\omega, \omega'_{2}).$ The weaker $\omega'_{i}$ recursion operators
		are given by:
		\begin{align}
		\widetilde{ T}'_{1} & :=  \omega^{-1} \circ \omega'_{1}  \cr
		& = \dfrac{\partial}{\partial J_{1}}\otimes dJ_{1} + \dfrac{\partial}{\partial \varphi^{1}}\otimes d\varphi^{1} - (2K + 1) \bigg( \dfrac{\partial}{\partial J_{2}}\otimes dJ_{2} 
		+  \dfrac{\partial}{\partial \varphi^{2}}\otimes d\varphi^{2}\bigg),
		\end{align}
		and
		\begin{align}
		\widetilde{T}'_{2} & := \omega^{-1} \circ \omega'_{2}  \cr
		& =  (2K + 1) \bigg(  \dfrac{\partial}{\partial J_{1}}\otimes dJ_{2} + \dfrac{\partial}{\partial \varphi^{2}}\otimes d\varphi^{1} \bigg)- \dfrac{\partial}{\partial \varphi^{1}}\otimes d\varphi^{2} - \dfrac{\partial}{\partial J_{2}}\otimes dJ_{1},
		\end{align}
		where \\
		$K =\dfrac{ (J_{1} + J_{2})^{3} }{m^{2}k^{2}\alpha}$ and the
		$2$-vector field
		$ \omega^{-1} = \dfrac{\partial}{\partial J_{i}}\wedge
		\dfrac{\partial}{\partial \varphi^{i} }.$
	\end{proposition}
	Similarly, $\omega''$ can be re-expressed as the sum of two $2$-forms
	as follows:
	\begin{equation}
	\omega'' =  \omega''_{1} +  \omega''_{2},
	\end{equation}
	where
	\begin{align}
	\omega''_{1} &= 2 dJ_{1}\wedge d\varphi^{1} - \bigg( 2J_{2} +  \dfrac{J_{2}(2J_{2} + 1)}{(J_{1} + J_{2})^{2}} \bigg)dJ_{2}\wedge d\varphi^{2} \\ 
	\omega''_{2} &=- J_{2}dJ_{1}\wedge d\varphi^{2}  - 2\dfrac{(1 + J_{1})\varphi^{1}}{(J_{1} + J_{2})} dJ_{1}\wedge dJ_{2}   + \bigg( 2  + \dfrac{2J_{2} + 1}{(J_{1} + J_{2})}\bigg)dJ_{2}\wedge d\varphi^{1} \cr 
	& + \bigg( \dfrac{2}{(J_{1} + J_{2})} + \dfrac{2J_{2} + 1}{(J_{1} + J_{2})^{2}}\bigg)(J_{2} - 1)\varphi^{1}dJ_{2}\wedge dJ_{1} . 
	\end{align}
	As above:
	
	\begin{itemize}
		\item[(iv)] $\omega''_{1}$ and $\omega''_{2}$ are not symplectic,
		$i.e.,$ $d\omega''_{1} \neq 0, d\omega''_{2} \neq 0$.
		\item[(v)]
		$ \iota_{_{X_{H}}}\omega''_{1} = -dh''_{1}, \quad
		\iota_{_{X_{H}}}\omega''_{2} = -dh''_{2}, $ where
		\begin{align}
		h''_{1} &= \dfrac{k^{2}m (3 J_{2}(8J_{2} - 6J_{1} + 3) + J_{1}(2J_{1} + 5))}{6(J_{1} + J_{2})^{3}}, \\
		h''_{2} &= \dfrac{ k^{2}m(J_{2}(-J_{2} - 3J_{1} + 12) + 8 J_{1})}{6(J_{1} + J_{2})^{3}}.
		\end{align}
		\item[(vi)] $h''_{1}$ and $h''_{2}$ are also first integrals of
		$X_{H},$ $i.e., X_{H}( h''_{1}) = X_{H}( h''_{2}) = 0.$
	\end{itemize}
	\begin{proposition}
		The Hamiltonian vector field $X_{H}$ is quasi-bi-Hamiltonian with
		respect to the two $2$-forms $(\omega, \omega''_{1}).$ Idem for
		$(\omega, \omega''_{2}).$ The weaker $\omega''_{i}$ recursion
		operators $\widetilde{T}''_{1}$ and $\widetilde{T}''_{2}$ are:
		\begin{align}
		\widetilde{T}''_{1} & := \omega^{-1} \circ \omega''_{1}  \cr
		& = 2\bigg(  \dfrac{\partial}{\partial \varphi^{1}}\otimes d\varphi^{1} + \dfrac{\partial}{\partial J_{1}}\otimes dJ_{1}\bigg) - J_{2}\bigg(2 + \dfrac{\widetilde{V}}{V^{2}} \bigg) \bigg( \dfrac{\partial}{\partial J_{2}}\otimes dJ_{2} +
		\dfrac{\partial}{\partial \varphi^{2}}\otimes d\varphi^{2}\bigg), \cr
		\\   
		\widetilde{T}''_{2} & := \omega^{-1} \circ \omega''_{2}  \cr
		& = \bigg( 2 +  \dfrac{\widetilde{V}}{V}\bigg) \bigg(  \dfrac{\partial}{\partial J_{1}}\otimes dJ_{2} + \dfrac{\partial}{\partial \varphi^{2}}\otimes d\varphi^{1} \bigg) -
		J_{2} \bigg(  \dfrac{\partial}{\partial \varphi^{1}}\otimes d\varphi^{2} + J_{2}\dfrac{\partial}{\partial J_{2}}\otimes dJ_{1} \bigg)\cr
		& -\bigg(2 + \dfrac{\widetilde{V}}{V^{2}} (J_{2} - 1) \bigg)\varphi^{1} \bigg( \dfrac{\partial}{\partial \varphi^{1}}\otimes dJ_{2} - \dfrac{\partial}{\partial \varphi^{2}}\otimes dJ_{1}\bigg),
		\end{align}
		where $\widetilde{V} = 2 J_{2} +1, \ V = J_{1} + J_{2}.$
	\end{proposition}
	
	\section{Concluding remarks} \label{sec_9} In this paper, we have
	constructed recursion operators for the Kepler dynamics in a deformed
	phase space by considering the equatorial orbit, computed the
	associated integrals of motion, and proved the existence of
	quasi-bi-Hamiltonian structures for the Kepler dynamics.
	
	\subsection*{Acknowledgments}	
	This work is supported by TWAS Research Grant RGA No.17-542 RG/
	MATHS/AF/AC\_G-FR3240300147.  The ICMPA-UNESCO Chair is in partnership
	with the Association pour la Promotion Scientifique de l'Afrique
	(APSA), France, and Daniel Iagolnitzer Foundation (DIF), France,
	supporting the development of mathematical physics in Africa.

\end{document}